# Methodological Advances and Challenges in Indirect Treatment Comparisons: A Review of International Guidelines and HAS TC Case Studies

Matthias Monnereau, MSc[1,2], Ana Jarne, PhD[2], Axel Benoist, PharmD[3], Clémence Fradet, PharmD[3], Maurice Perol, MD, PhD[4], Thomas Filleron, PhD[1,5] *, Louise Baschet, MSc[2] *

* Thomas Filleron and Louise Baschet contributed equally to this work

1. Université Paris-Saclay, CESP, INSERM U1018 Oncostat, labeled Ligue Contre le Cancer, Villejuif, France
2. Horiana, Bordeaux, France
3. Astra Zeneca, Courbevoie, France
4. Department of Medical Oncology, Léon Bérard Cancer Centre, Lyon, France
5. Biostatistics & Health Data Science. Oncopole Claudius Regaud – IUCT-Oncopole. 1 avenue Irène Joliot Curie, Toulouse, France

**Corresponding authors:**

Thomas Filleron,
Oncopole Claudius Regaud, IUCT-Oncopole
1 avenue Irène Joliot Curie
31059 Toulouse Cedex, France
Filleron.thomas@iuct-oncopole.fr

**Co-Corresponding author:**

Louise Baschet
Horiana
80bis rue Paul Camelle
33100 Bordeaux, France
Louise.Baschet@horiana.com

**Financial disclosure**: This study was supported by Astra Zeneca

**Acknowledgements**: We would like to thank Mathilde Grande, PharmD, Mathieu Robain, MD PhD, and Michel Cucherat, MD PhD, for their constructive comment in data interpretation and manuscript review.

# ABSTRACT

To evaluate methodological challenges and regulatory considerations of indirect treatment comparisons (ITCs) with the analysis of international health technology assessment guidelines and French Transparency Committee (TC) decisions. We conducted a pragmatic review of ITC guidelines from major health technology assessment (HTA) bodies and multistakeholder organizations. Then, we analyzed TC opinions published between 2021-2023. We extracted data on ITC methodology, therapeutic areas, acceptability, and limitations expressed by the TC. The targeted review of the main guidelines showed mainly agreements between HTA bodies and multistakeholder organizations, with some specificities. 138 TC opinions containing 195 ITCs were analyzed. Only 13.3% of these ITCs influenced TC decision-making. ITCs were more frequently accepted in genetic diseases (34.4%) compared to oncology (10.0%) and autoimmune diseases (11.1%). Methods using individual patient data showed higher acceptance rates (23.1%) than network meta-analyses (4.2%). Main limitations included heterogeneity/bias risk (59%), lack of data (48%), statistical methodology issues (29%), study design concerns (27%), small sample size (25%), and outcome definition variability (20%). When ITCs were the primary source of evidence, the proportion of important clinical benefit was lower (60.9% vs. 73.4%) than when randomized controlled trials were available. While ITCs are increasingly submitted, particularly where direct evidence is impractical, their influence on reimbursement decisions remains limited. There is a need for clear and accessible guides so manufacturers can produce clearer and more robust ITCs that follow regulatory guidelines, from the planning phase to execution.

# INTRODUCTION

The use of indirect treatment comparisons (ITCs) as evidence for assessing a treatment's efficacy and safety in reimbursement processes has substantially increased in recent years (1,2), driven by specific needs in health technology assessment :

- The need to compare a new treatment to comparators different from those used in pivotal trials
- The need to obtain comparative results in the context of non-comparative studies

These situations reflect the increasing complexity of treatment pathways, specifically:

- Concurrent development of several innovative treatments for the same condition
- Growing number of studies focusing on rare diseases, where non-comparative studies are often conducted
- Situations where comparator choices differ between regions of the world or where suboptimal comparators have been used
- Possibility for early access to innovative treatments while RCTs are ongoing or pending

The proliferation of clinical databases and health registries, along with recent FDA encouragement for rapid access to innovative treatments (3,4), has resulted in a vast amount of real-world data (RWD), facilitating the use of ITCs. However, despite the enthusiasm for these techniques, several methodological issues inherent to the indirect nature of these comparisons remain. As such, the different regulatory agencies and health technology assessment (HTA) bodies have published guidelines on the appropriate use of ITCs.

The statistical methods used in ITCs can be put into different categories:

When several randomized clinical trials (RCT) exist in the targeted population:

- Network meta-analysis (NMA) with the Bucher method or Bayesian mixed treatments comparisons (MTC),
- Multi-level network meta-regression (ML-NMR)
- Propensity score

In the absence of a common comparator:

- Propensity score (adjustment, matching, weighting, stratification),
- Matching Adjusted Indirect Comparisons (MAIC)
- Simulated Treatment Comparisons (STC)

This paper aims to describe reimbursement submissions containing ITCs from the Transparency Committee (TC) with regard to HTA bodies and multistakeholder organizations guidelines.

# METHODS

This article are based on two complementary works; a pragmatic review of various guidelines on indirect treatment comparisons (both French and international), and a systematic review and analysis of all opinions published by the TC from January 1$^{st}$ 2021 to December 31$^{st}$ 2023 that included at least one indirect comparison.

## *Guidelines review*

We conducted a pragmatic review on ITC methodology and implementation in clinical trials. We screened HTA bodies' websites[1] with the following search sentences: *indirect treatment comparison guidelines/guidance*, *network meta-analysis guidelines/guidance*, *adjusted indirect comparison guidelines/guidance*, and *external/synthetic control guidelines/guidance*. Additionally, we conducted a targeted search on Pubmed (to see Supplementary methods), and used Google to capture HTA guidelines not present on official websites or for organizations without dedicated websites, applying the same search sentences.

---

[1] https://www.has-sante.fr/, https://www.fda.gov/, https://www.nice.org.uk/, https://www.iqwig.de/en/, https://www.ema.europa.eu/en, https://www.eunethta.eu/, https://www.ispor.org/.

## *Analysis of opinions published by the TC of the HAS*

We analyzed all ITCs mentioned in TC dossiers between January 1, 2021, and December 31, 2023, and reviewed the HAS's consideration of these methods. An automated algorithm based on key words screened the opinions of the HAS for treatments using ITCs (see Supplementary Materials). Each opinion was manually reviewed, and each distinct ITC presented in the opinion was extracted. A second reviewer checked 10% of the opinions for quality assurance. Both qualitative analysis and quantitative descriptive analysis were conducted, the unit was the ITC, as multiple ITCs could appear in a single opinion.

A data extraction template was developed in Microsoft excel. There were 41 variables either extracted automatically via the website's Application Programming Interface (API) or manually.

Variables can be specific to the ITC(s) presented in the opinion: type of indirect comparison, source of clinical evidence (RCT-based or single-arm), acceptability of the ITC, main source (i.e., primary source submitted by the industry to claim reimbursement for the drug, e.g., in absence of an RCT) for comparative effectiveness and more. They can also be related to the opinion itself: name of the product, clinical benefit (CB, *Service Médical Rendu* or SMR), clinical added value (CAV, *Amélioration du Service Médical Rendu* or ASMR) granted by the TC or requested by the manufacturer and more. The SMR/CB can be either Important, Moderate, Low or Insufficient. No ASMR/CAV is given if an SMR/CB insufficient is obtained. All variables extracted can be found in Supp Mat 1.

In addition to the 41 variables, the limitations related to each ITC in the opinion bought up by the TC were also extracted and interpreted. They were divided into 3 main categories: Limitations related to the data, to the methodology and to uncertainty. Those categories were then divided into subcategories and themselves divided into other smaller categories to achieve a satisfactory level of detail. All categories can be found in Supp Mat 2.

The overall acceptability of the ITC was defined based on the use of the results of the ITC (and its methodology) by the TC for wording of the ASMR. The decision-making of the ASMR appears in the summary part of each opinion made by the TC, where justifications for each decision is given. If the ITC was mentioned to not be acceptable or not mentioned at all in the decision-making of the ASMR, then it was categorized as "Not acceptable". If it was considered for the decision-making but with clear stated issues, it was categorized as "Not rejected". If it was considered and no issues were clearly stated, it was categorized as "Acceptable". Others were categorized as "Not applicable / Not mentioned / Unclear" depending on the situation. As such, the acceptability of the ITC is not based upon the overall reimbursement of the drug but rather upon the part it played on the decision-making of the ASMR.

# RESULTS

## *Guidelines review*

A total of 29 documents covering the recommendations on this topic of 7 different bodies were included, which are: recommendations from the HAS in 2009 (5), 2020 (6), 2021 (7) and 2023 (8,9), the European Medicines Agency (EMA) in 2007 (10), 2020 (11) and 2023 (12), the EUropean network for Health Technology Assessment (EUnetHTA) in 2022 (13,14) that have been endorsed by the HTA CG in 2024 (15,16), the National Institute for Health and Care Excellence (NICE) between 2008 and 2022 (17–28), the Food and Drug Administration (FDA) in 2018 (29) and 2023 (30), the International Society for Pharmacoeconomics and Outcomes Research (ISPOR) in 2011 (31,32), 2014(33) and 2017 (34), and the *Institut für Qualität und Wirtschaftlichkeit im Gesundheitswesen* (IQWiG) in 2023 (35). These guidelines emerged to address the challenges arising from the increasing use of indirect comparison methods. All recommendations found in the guidelines were listed and synthesized, then cross-checked to show a summary of all recommendations by agencies (see table 1). The proper conduct of a systematic literature review is an important part of the recommendations from HAS, NICE and IQWiG. All three HTA bodies agree on emphasizing the importance of a clear and complete trials selection, a priori identification of potential confounders, prognostic factors and biases (depending on whether the situation is anchored or unanchored and on the method used), and a correct reporting of the heterogeneity, inconsistency, and concordance with previous findings and literature.

Focusing on Network Meta-analyses, the most frequently repeated recommendations are the homogeneity and consistency within clinical trials, the justifications of clinical assumptions and methodological choices.

Regarding External Controls Arms (ECA) using IPD, identifying and properly managing potential confounders is universally agreed upon, but there are also other widely accepted criteria, like anticipating the construction of the ECA or conducting sensitivity analyses.

**Table 1. Comparative Analysis of Guidelines on Indirect Comparisons Design**

| | HAS | NICE | IQWiG | EMA | FDA | ISPOR | EUNetHTA |
|---|---|---|---|---|---|---|---|
| **Systematic Literature Review** | | | | | | | |
| Exhaustive systematic review with critical appraisal | X | X | X | - | - | - | - |
| Clear and complete trials selection criteria | X | X | X | - | - | - | - |
| Avoiding publication bias | X | X | X | - | - | - | - |
| A priori identification of potentially relevant confounders, effect modifiers, prognostic variables or biases, when appropriate | X | X | X | - | - | - | - |
| Reporting heterogeneity, external validity, inconsistency | X | X | X | - | - | - | - |
| Clear conclusions | X | X | X | - | - | - | - |
| Define the research question | X | X | X | - | - | - | - |
| At least 2 high-publications | O | X | X | - | - | - | - |
| Quality assessment according to guidelines (10) or (11,12) | O | O | X | - | - | - | - |
| **Network Meta Analysis** | | | | | | | |
| Homogeneity of the clinical trials characteristics | X | X | X | - | X | O | X |
| Consistency | X | X | X | - | O | X | X |
| Diagnosis of convergence if bayesian approaches | X | X | O | - | O | O | X |
| Justification of clinical assumptions and methodological choices | X | X | X | - | O | X | X |
| Principles of good practice for standard pairwise meta-analyses should also be followed in adjusted indirect treatment comparisons and network meta-analyses | X | X | X | - | O | O | O |
| Respect within-study randomization | O | X | O | - | O | X | X |
| Complete description of the model | X | X | X | - | O | O | X |
| Ideally, the network meta-analysis should contain all identified treatments | X | X | O | - | O | X | X |
| Targeted towards the overall research question of interest, not only towards selective components such as individual outcomes | O | O | X | - | O | O | O |
| Follow their consensus-based 26 item questionnaire | O | O | O | - | O | X | O |
| Describe sensitivity analyses, if remaining unclear issues | X | X | X | - | O | O | X |
| **External Control Arm using Individual Patients Data** | | | | | | | |
| Anticipated and scheduled in advance | X | X | O | X | X | - | O |
| Performing transparent and appropriate analyses | X | X | O | X | O | - | X |
| Causal inference analysis: matching, weighting, regression, … | X | O | O | O | O | - | O |
| Potential bias considerations | X | X | O | X | X | - | X |
| Identify potential confounders, clearly articulate assumptions, use a statistical method that considers observed and unmeasured confounders, | X | X | X | X | X | - | X |
| Sensitivity analysis | O | X | X | X | X | - | O |
| Justify the need for non-randomised evidence | X | X | O | X | O | - | O |
| Design studies to emulate the target randomised controlled trial | X | X | O | O | O | - | O |
| Explore goodness of fit | X | O | X | O | O | - | O |
| Handle missing data | X | X | X | X | X | - | O |
| **Population Adjusted Indirect Comparisons** | | | | | | | |

| | HAS | NICE | IQWiG | EMA | FDA | ISPOR | EUnetHTA |
|---|---|---|---|---|---|---|---|
| Justify the use against standard ITC | X | X | O | O | - | - | O |
| Discuss consistency and relevance of the pseudo-population | X | X | O | O | - | - | O |
| Prefer use of Individual Patients Data in all sources | X | O | X | O | - | - | X |
| Prefer anchored forms if possible | O | X | O | O | - | - | X |
| Provide evidence on the likely extent of error due to unaccounted for covariates, especially in case of unanchored forms | O | X | O | O | - | - | X |
| Define target population, and appropriate measure of uncertainty | X | X | O | O | - | - | O |
| Select the covariables | O | X | O | O | - | - | O |
| Warms on limitations and advices to justify | X | X | O | X | - | - | X |
| Transparent report of identification of all relevant variables, prognostic variables and effect modifiers. ML-NMR as an alternative to Network Meta-Analysis | O | O | O | O | - | - | X |
| **Decision tree to guide the process of indirect comparisons** | | | | | | | |
| Basic diagram for indirect comparison justification | X | O | - | - | - | - | - |
| **Flowchart to select a method for indirect comparisons** | O | X | - | - | - | - | - |

X: this recommendation appears in guidelines from the corresponding institution; o: this recommendation does not explicitly appear in guidelines from the corresponding institution. -: this section is not mentioned at all in guidelines from the corresponding institution

HAS: Haute Autorité de la Santé; NICE: National Institute for Health and Care Excellence; IQWiG: Institute for Quality and Efficiency in Health Care; EMA: European Medicines Agency; FDA: Food and Drug Administration; ISPOR: International Society for Pharmacoeconomics and Outcomes Research; EUnetHTA: European Network for Health Technology Assessment.

As for the Population-Adjusted Indirect Comparisons (PAIC), the most common recommendation is to discuss limitations of the implemented methods and provide advice for justification. Another criterion is the justification of the use of an ITC using aggregated data against a more "standard" ITC such as NMA or methods using IPD, and to discuss consistency and relevance of the pseudo-population. An important note is that unanchored PAICs do not possess the same weight of evidence as anchored PAICs because of the more restrictive hypotheses inherent to their nature (no unmeasured prognostic factors). As such, while this recommendation stands for all ITCs and echoes to SLR recommendations, the highest care must be used for implementing all effect modifiers and prognostic variables and provide clear justification and evidence on the likely extent of error due to unaccounted covariates in relation to the observed relative treatment effect. An exception can be made for IQWIG as it strongly advise against using unanchored PAICs.

Overall, while specific recommendations can be observed by country, it can be concluded that these guidelines do not present substantial differences that would challenge the acceptability of a unified approach within the EU, nor justify that global operations must systematically prepare separate documentations. However, given that decisions are still made at the national level, the focus of the paper from this point onward will remain on France with the Transparency Committee of the HAS.

## Analysis of opinions published by the TC of the HAS

In total, 1,218 opinions were published by the TC between 2021 and 2023, of which 138 (11.3%) contained at least one indirect comparison and were systematically screened. The number of opinions containing ITCs increased by 44.7% between 2021 (n=38) and 2023 (n=55). In these 138 opinions, 195 ITCs were identified (of which, 46.7% were the primary source of evidence for the comparison).

Among the 138 opinions published by the Transparency Committee containing at least one indirect comparison, most are either for initial marketing authorizations (32.6%) or for indication extensions (32.6%), but there are also early access authorizations (19.6%) and requests for reevaluation (15.2%). Half of the opinions concern oncological diseases (50.7%). In 15.9% of cases, they relate to genetic diseases, and 10.1% target autoimmune diseases. Also, 17.4% of the opinions concern pediatric diseases, and 16.7% concern rare diseases (see table 2).

| Table 2. Main characteristics of the 138 identified opinions | | |
|---|---|---|
| Procedure | Initial assessment | 45 (32.6%) |
| | Indication extension | 45 (32.6%) |
| | Early access authorization | 27 (19.6%) |
| | Reassessment | 21 (15.2%) |
| Disease area | Oncology | 70 (50.7%) |
| | Genetic disease | 22 (15.9%) |
| | Autoimmune diseases | 14 (10.1%) |
| | Other | 32 (23.2%) |
| Additionnal information | Pediatric diseases | 24 (17.4%) |
| | Rare diseases | 23 (16.7%) |

Among the 195 indirect comparisons, 26 (13.3%) were considered by the TC (one can assume that the results of the indirect comparison were considered in the decision-making process) for the opinion assessment, and 169 (86.7%) were not considered. Regarding the rationale for conducting the ITC, while not explicitly stated in half of the cases, it involves the presence of multiple comparators in 16.9% of cases, the concomitant development of two "competing" drugs in 14.4% of cases, or the need to compare against real-world data in 10.8% of case (see table 3).

| Table 3. Characteristics of the 195 identified ITCs | | |
|---|---|---|
| Rationale for ITC provided by the manufacturer | Several comparators | 33 (16.9%) |
| | Concomitant development | 28 (14.4%) |
| | Need to compare to SoC/RWD | 21 (10.8%) |
| | Mandatory from HTA | 1 (0.5%) |
| | Non-feasibility of the RCT | 2 (1.0%) |
| | Not mentioned | 110 (56.4%) |
| ITC predefined in the protocol | Yes | 70 (36.6%) |
| | No | 16 (8.4%) |
| | Not mentionned | 54 (28.3%) |
| | Unclear | 51 (26.7%) |
| ITC mentioned in the TC summary | No | 109 (55.9%) |
| | Yes | 74 (37.9%) |
| | Unclear | 12 (6.2%) |

*SoC: Standard of Care; RWD: Real World Data; RCT : Randomized Control Trial

If the ITC is the main source for comparing effectiveness, the percentage of important SMR obtained is lower (60.9%) than when an RCT is the main source of comparison (73.4%), while the percentage of insufficient SMR is higher (18.8% vs. 9.6%). Additional information can be found in Supp Mat 3. It is difficult to conclude on the correlation between the main source of comparison and the obtained ASMR. 3% of ITCs were accepted whether they were the main source of evidence (3/92) or not (3/103). However, 3% (3/103) of ITCs who were not the main source of evidence were not rejected compared to 19% (17/92) for those who were (see figure 1).

**Figure 1. ITCs served as the main source (i.e., primary source submitted by the industry to claim reimbursement for the drug, e.g., in absence of an RCT) for comparing effectiveness and acceptability**

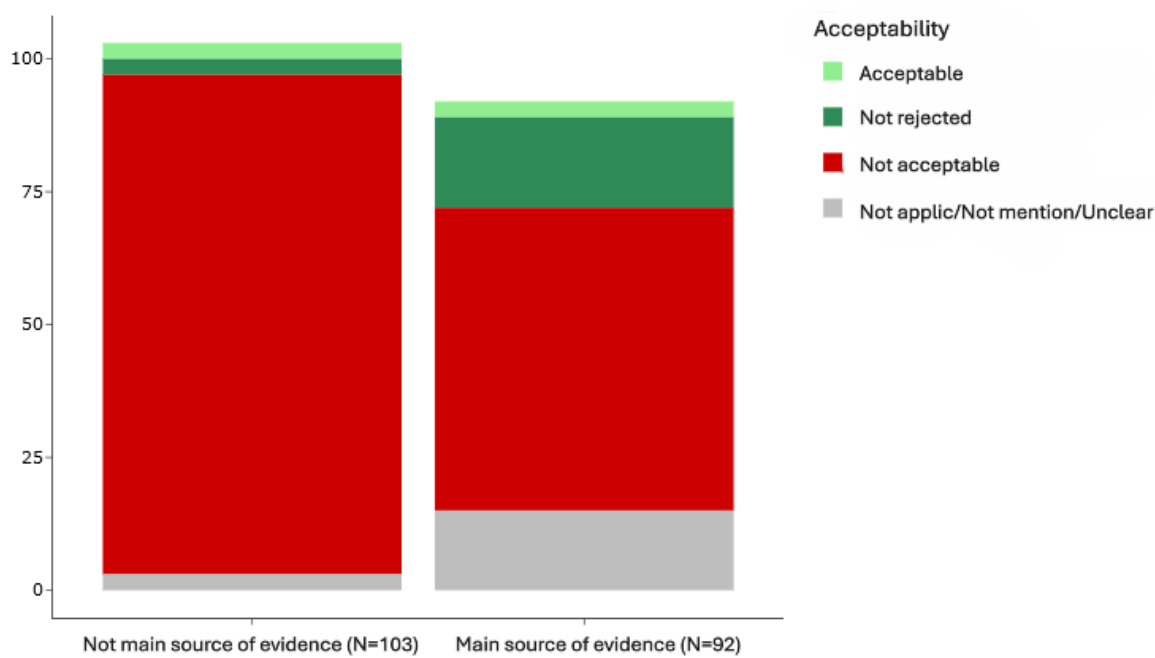

Looking at the different therapeutic areas, the percentage of acceptance of indirect comparisons in opinions concerning genetic diseases (n = 32) was 34.4%, while this percentage was 11.1%, 10.0%, and 6.7% for autoimmune diseases (n = 18), oncological diseases (n = 100), and others (n = 44), respectively. In this line, it is observed that ITCs conducted in the context of pediatric diseases and rare diseases have a higher acceptance rate. Also, the percentage of accepted indirect comparisons is higher when the ITC is the main source for comparing effectiveness (21.6% vs 5.8%).

The highest percentage of accepted and/or not rejected comparisons is found among unadjusted comparisons (3/8: 37.5%). They are followed by Inverse Probability Treatment Weighting / Standardized mortality ratio weighting, Matching & Other with IPD methods (9/39: 23.1%); Matching-Adjusted Indirect Comparisons (MAIC), Simulated Treatment Comparisons (STC) based on Single-arm (8/36: 22.2%), Other/not defined (2/11: 18.2%), MAIC / STC based on RCT (1/13: 7.7%), and Network Meta Analysis (NMA) / Bucher (3/71: 4.2%) (see figure 2).

# Figure 2. Type of ITC and acceptability

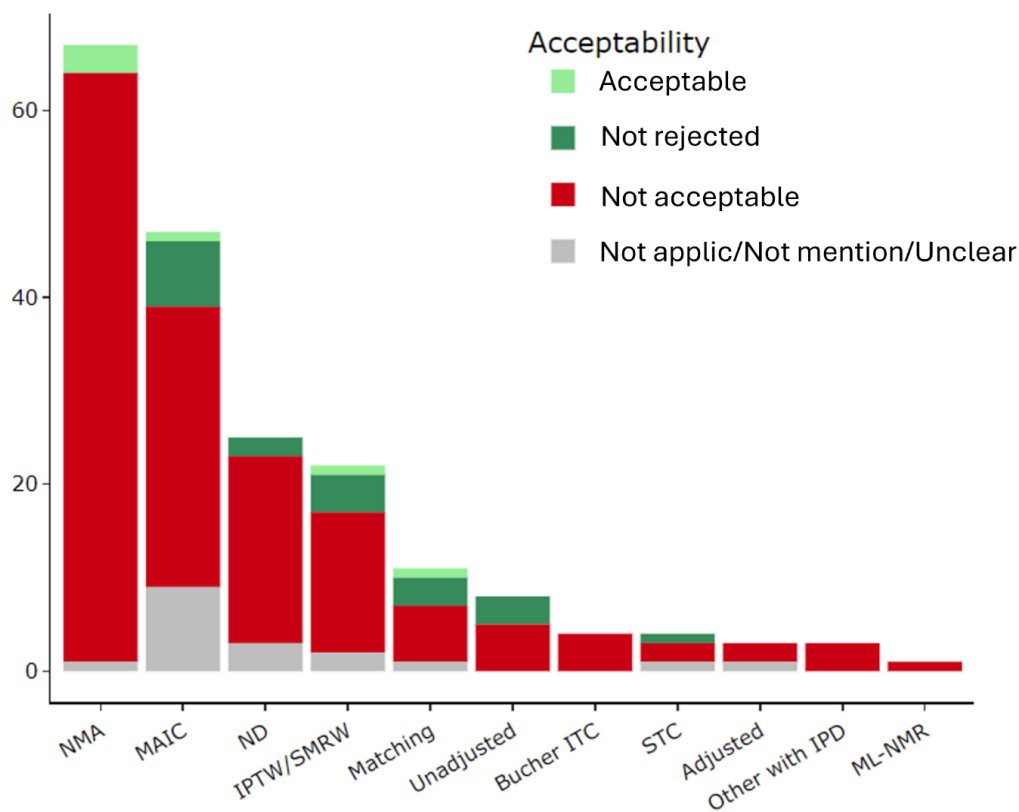

A total of 6 major limitations were found in the 195 ITCs with a frequency of over 20%: "Heterogeneity / risk of bias" (59%), a "lack of data / unclear data" (48%), "limitations in the statistical methods" (29%), "Study design" (27%), "Sample size" (25%) and "Variability in definition/timing of the outcomes" (20%) (See Table 4 for more frequently encountered limitations and Supp Mat 3 for limitations by class of statistical methods used).

| Table 4. Most frequently encountered limitations | |
|---|---|
| Heterogeneity / risk of bias | 59% |
| Lack of data / unclear data | 48% |
| Limitations in the statistical methods | 29% |
| Study design | 27% |
| Sample size | 25% |
| Variability in definition/timing of the outcomes | 20% |
| Choice of comparators | 12% |
| Proper tracing of feasibility | 8% |
| Post hoc decision on indirect comparison | 5% |
| Insufficient justification | 2% |
| Lack of sensitivity analyses | 2% |

# DISCUSSION

The analysis of opinions published by the TC of the HAS, alongside our review of international guidelines, reveals important insights into the theoretical framework and practical application of ITCs in HTA decision-making.

Overall, 11.3% of TC opinions (n=138) included at least one ITC (n=195) in a span of 3 years (2021-2023) with an increase of 44.7% between 2021 (n=38) and 2023 (n=55), showing that these methods, while still new, are beginning to flourish and are considered by manufacturers to inform decision-making of the value assessment of their treatment. However, only 13.3% (26/195) of these ITCs were actually taken into account by the TC in its decision-making. This gap suggests that ITCs don't often impact the final outcomes of reimbursement, even when they are the sole evidence of comparative effectiveness (46.7% of ITCs).

The comparative analysis of ITC guidelines across multiple HTA bodies and regulatory agencies revealed shared guidelines and divergences. As shown in Table 1, all three major HTA bodies (HAS, NICE, and IQWiG) demonstrate strong alignment on systematic literature review requirements, including exhaustive reviews with critical appraisal, clear trial selection criteria, and a priori identification of potential confounders. However, methodological variations emerge across agencies in other domains. For NMAs, while homogeneity of clinical trials and consistency are important considerations by most bodies, diagnostic approaches for convergence in Bayesian methods was not the case, with HAS and NICE providing guidance while the FDA, IQWIG and ISPOR offer limited direction. Similarly, for ECAs using IPD, there is an agreement on the importance of potential bias considerations and confounder identification, but differences exist regarding sensitivity analysis requirements, emphasized by NICE, HAS, FDA, and EMA but not explicitly mentioned ISPOR and EUNetHTA guidelines. The most pronounced differences appears in PAICs, where agencies like EUnetHTA and HAS strongly recommend the use of IPD instead of PAICs in all data sources while others provide more explicit guidance and IQWIG having a strict stance against the use of PAICs. These nuanced differences in methodological emphasis may partially explain the varying acceptance rates of different ITC types observed in TC submissions.

The lower acceptance of ITCs in oncological and autoimmune diseases (11.1% and 10.0%, respectively) is particularly noteworthy. One reason could be that these areas often benefit from strong direct evidence with randomized controlled trials (RCTs), making ITCs less critical. This is in line with HTA recommendations, which advise prioritizing direct evidence and using ITCs only when head-to-head comparisons aren't feasible (HAS, 2019). Conversely, ITCs were more commonly considered in genetic diseases, pediatric conditions, and rare diseases. This aligns with recommendations for using ITCs in areas with significant unmet medical needs, where direct evidence is often scarce due to small patient populations. The higher reliance on ITCs in these cases reflects the flexibility of HTA bodies in accepting alternative evidence-generation methods when traditional approaches are impractical (IQWiG, 2020). However, HTA guidelines all emphasize the importance of addressing potential biases and confounding factors, particularly in small or heterogeneous populations, where ITCs can offer valuable insights despite their limitations.

For the reasons just mentioned, indirect comparisons sometimes occur in contexts where a very small number of patients may have benefited from an extraordinarily effective treatment. This could explain why unadjusted comparisons have the highest considered rate between all other methods, despite their methodological flaws. Adjustment methods using IPD are the second highest accepted methods (23.1%). This is in accordance with the different guidelines as IPD methods are considered by HTA bodies as having a stronger evidence base compared to methods using aggregated data, mostly because they take full advantage of all patient characteristics. However, when looking at methods using aggregated data, unanchored forms (comparisons with a single arm) have a higher consideration than anchored forms (comparisons with an RCT), with 22.2% and 7.7% respectively. While this difference can be partially explained by the difference in the number of cases analyzed (36 for unanchored and 13 for anchored), heterogeneity remains between the acceptance of both forms of methods. Anchored forms of adjustments are held in higher regards than unanchored forms by HTA bodies as they add more hypothesis difficult to validate. This could suggest that when an industrial submits an unanchored indirect comparison using aggregated data, it is because no network could be established and that the medical need or at least the evidence base for this pathology is weak, allowing for more flexibility in the reviewing process. Additionally, the limited consideration of NMAs and Bucher ITC at just 4.2% when it was the most often found comparisons (n=71) might suggest that the very strict criteria for homogeneity and consistency across clinical trials are not always met.

One important finding is that when ITCs were the primary source of evidence (typically when RCTs versus comparators were impossible to conduct), the proportion of significant SMR was lower (60.9% vs. 73.4%), and the proportion of insufficient SMR was higher (18.8% vs. 9.6%). In cases where justification for conducting an ITC was provided, it was predominantly due to the impossibility of conducting an RCT against the chosen comparator. This observation reinforces that RCTs remain the most positively assessed type of study in terms of clinical evidence. While this preference for RCT evidence is logical, it highlights important challenges for the access of innovative treatments in France today, particularly in situations where ITCs are the only feasible option.

## CONCLUSION

In summary, while the different guidelines provide a strong foundation for the use of ITCs, their practical application in France demonstrates both the usefulness and limitations of these methods. ITCs are more frequently considered in therapeutic areas with high unmet medical needs, such as genetic and rare diseases, where direct evidence is often lacking. However, their influence on decision-making remains limited in areas where robust direct evidence is available, such as oncology and autoimmune diseases. This emphasizes the need for clear and accessible guides to the attention of manufacturers to make clearer and more robust ITCs that follows as closely as possible HTA guidelines, from planification of the ITC to the application of precise statistical methodologies and appropriate sensitivity analysis.

# SUPPLEMENTARY MATERIAL

**Appendix 1 – Variables extracted**

- Common by opinion:
    - Automatically extracted:
        - Review ID
        - Evamed Number
        - link,
        - Product
        - Categories
        - Type of procedure
        - Summary
        - Nature of the procedure
        - SMR
        - ASMR
        - Manufacturer
        - Priority notice (YES/NO)
        - Early Access
        - Validation date
    - Manually extracted
        - Area of therapy/pathology
        - Sector (hospitall/ primary care/ hospital and primary care)
        - Indication details (rare disease/ orphan treatment)
        - SMR and ASMR requested by the manufacturer (information available in the transcription file)
        - Data package (i.e. list of submitted studies in the dossier)
        - Indication restriction in conclusion(Y/N)
- Specific by ITC:
    - Type of Indirect comparison (NMA, NMR, Bucher ITC, MAIC, STC, IPTW, Matching, Adjusted, Unadjusted, Other without IPD, Other using IPD, ND)
    - Details on the statistical methods (with copy-pasted information from the opinion)
    - Sources of clinical evidence used in ITC (RCT-based ITC, single-arm study-based ITC, other)
    - ITC as main evidence source (Y/N – cf. data package)
    - SMR and ASMR for the population of the ITC
    - If early access: authorization (refused/ granted/ renewed)
    - ITC data taken into account regarding the conclusions of the HAS ("Yes","No","Unclear")
    - ITC was published in a scientific paper

- Reason for ITC: non-ethical, non-feasible (rare disease), concomitant development, many comparators, other, not mentionned, need to compare to SoC/ RWD, mandatory from HTA ...)
- ITC predefined in the protocol/during design phase (a priori / a posteriori)
- Literature review for ITC data sources (Y/N, type of literature review, methodology)
- Literature review for confounders and effect modifiers (idem)
- Protocol, SAP and reports detailed enough to assess study
- A priori definition of population, comparators, outcomes of interest
- Model predefined with appropriate confounding
- Model based on individual data only
- Underlying assumptions of model explored and met
- If trimming (keeping a subset of population study), the resulting target population is described
- Residual confounding explored
- Efficacy results: conclusion (in favor, neutral ...), magnitude
- Safety results: conclusion (in favor, neutral, not reported ...), magnitude, reported for all populations (YES/NO)
- Critics of the ITC methods raised by the HAS
  - Limitations related to data:
    - weakness of the justification of ITC in place of RCT (YES/NO + details)
    - Study design weakness
      - due to the type of studies included,
      - due to the inclusion criteria of bibliographic search of data
      - details – page XX
    - Sample size
      - small patient number
      - small sample size after covariates matching,
      - small studies number,
      - uncertainty related to wide credible interval
      - details – page XX)
    - Interventions/comparators
      - Comparators different from SoC,
      - Other
      - details – page XX
    - Outcomes (variability of the definition or timing of the outcomes + details)
    - Heterogeneity/risk of bias
      - Heterogeneity or Inconsistency,
      - Inconsistencies in the study search and population not well described,
      - Impossible to assess heterogeneity / Lack of bias assessment,
      - Other
      - details – page XX
    - Lack of/unclear data
      - Few baseline's covariates to match patients,
      - Weak evidence/post-hoc data,

- o   Individual patient data not available,
- o   Lack of data on effect modifiers/ prognostic factors,
- o   Prognostic factors not justified,
- o   Data was not presented adequately,
- o   Missing validity of endpoints, quality of trials and evidence,
- o   Uncertainty in the source of data,
- o   Data not reported for some endpoints/lack of comparative data,
- o   Immature data,
- o   details – page XX
- Limitations related to methodology:
  - Anticipation in design phasis (ITC post-hoc +details)
  - Feasibility of the ITC (
    - o   Comparison does not fulfill the requirements of an adjusted indirect comparison,
    - o   Non comparable outcomes,
    - o   Other
    - o   +Details)
  - Statistical methods
    - o   Wrong choice of extrapolation curves,
    - o   Assumptions not met,
    - o   Uncertainty related to proportional hazard,
    - o   Method of adjustment/ limitation of the matching for prognostic factors,
    - o   Estimated models (fixed/random effect, fractional polynomials),
    - o   Adequate data for handling subgroups was missing,
    - o   Description for the simulated comparison was missing,
    - o   No hypothesis on the expected efficacy has been made,
    - o   +Details
- Limitations related to uncertainty:
  - Missing sensitivity analysis
  - Predictive validity versus real-world data
  - Details
- Other comments

Some columns in the extraction grid will be specific to certain methods:

- NMA: Number of studies, number of comparators
- STC /MAIC: ESS, comparators & external sources chosen independently of the results and fit clinical question, study characteristics of uncontrolled trial and comparator are similar enough, anchored/unanchored

**Appendix 2 – Categories and sub-categories of limitations**

| Main category (limitations related to…) | Sub-categories | Detailed categories |
|---|---|---|
| Related to data | Weakness of the ECA's justification | / |
| | Study design | Due to the type of studies used in the comparison |
| | | Due to the inclusion criteria of the bibliographic search and data used |
| | Sample size | Small patient number |
| | | Small sample size after covariates matching |
| | | Small number of studies |
| | | Uncertainty related to wide credible intervals |
| | Treatments/Comparators | Comparators different from Standard of Care (SoC) |
| | | Other |
| | Outcomes | Variability in the definition and timing of the outcomes |
| | Heterogeneity / Risk of bias | Heterogeneity/inconsistency |
| | | Inconsistencies in the study search/population not enough described |
| | | Impossible to assess heterogeneity/ Lack of bias assessment |
| | | Other |
| | Lack of/Unclear data | Few baseline covariates available |
| | | Weak evidence/post-hoc data |
| | | Individual patient data not available |
| | | Lack of data on effect modifiers/ prognostic factors |
| | | Prognostic factors not justified |
| | | Data was not presented adequately (lacking traceability/inadequate inclusion of studies ...) |
| | | Missing validity of endpoints, quality of trials and evidence |
| | | Uncertainty in the source of data |
| | | Data not reported for some endpoints; lack of comparative data (eg safety) |
| | | Immature data |
| | | Other |
| Related to methodology | Anticipation in the design phase: study designed for the ECA | ECA post hoc |
| | Feasibility of the ECA | Comparison does not fulfil the requirements of an adjusted indirect comparison |
| | | Non comparable outcomes |
| | | Other |
| | Statistical methods | Assumptions not met |
| | | Method of adjustment/ limitation of the matching for prognostic factors |
| | | No hypothesis on the expected efficacy has been made |
| | | Management of missing data |
| Related to the uncertainty | Missing sensitivity analysis | Lack of sensitivity analysis |
| | Predictive validity versus real world data | Treatment effect was tested on ideal population, the effect could change in real conditions |

# Appendix 3 – Frequency of limitations for each category of statistical method used

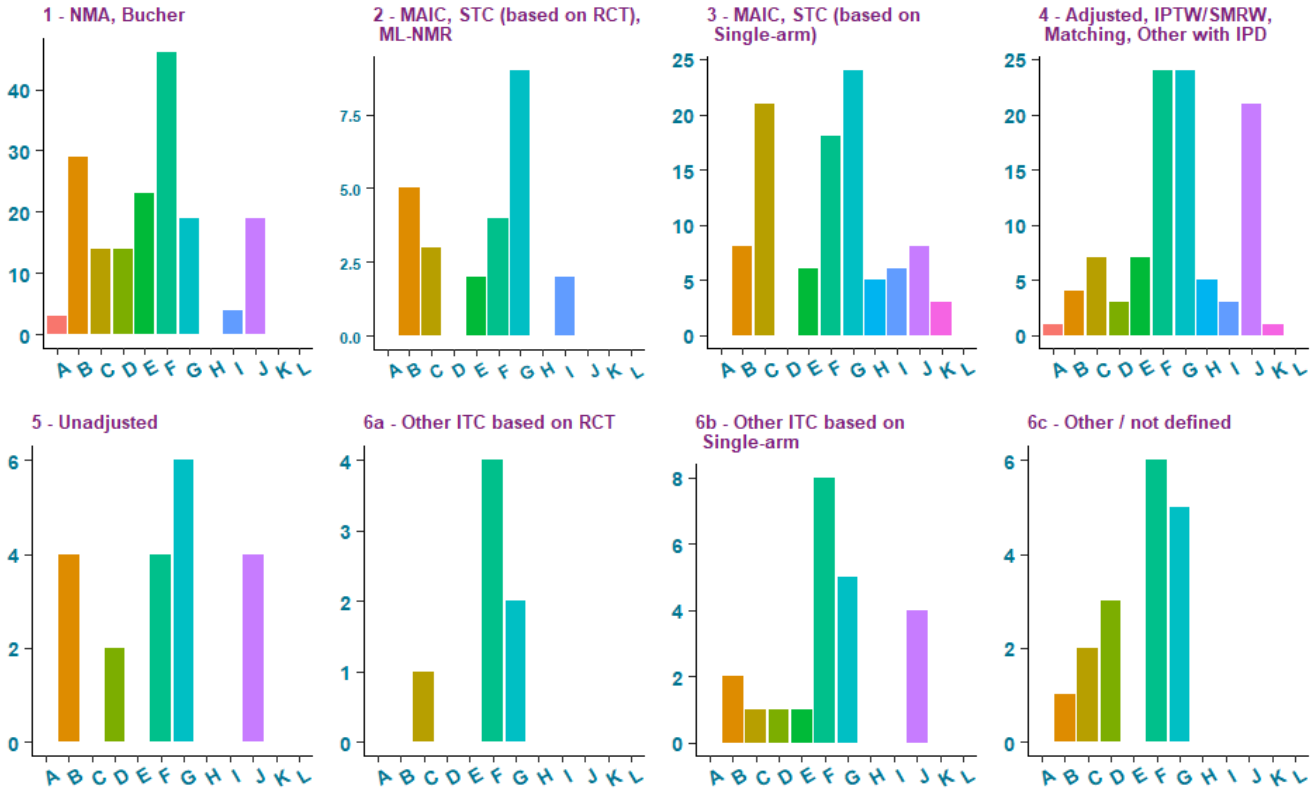

**A** - Weakness of justification of ITC ; **B** - Study design ; **C** - Sample size ; **D** - Treatment / Comparators **E** – Variability ; **F** - Heterogeneity / Risk of bias ; **G** - Lack of / Unclear data ; **H** - ITC post hoc ; **I** – Feasibility ; **J** - Statistical methods ; **K** - Lack of sensitivity ; **L** - Ideal population